\documentclass[12pt,a4]{article}
\usepackage{amsmath,amssymb}
\usepackage{epsfig,pstricks}

\newcommand{\im}{\mathrm i}
\newcommand{\C}{\mathbb C}
\newcommand{\tr}{\operatorname{Tr}}
\newcommand{\llg}{\operatorname{ln}}

\newcommand{\eq}{\begin{equation}}
\newcommand{\en}{\end{equation}}
\newcommand{\bear}{\begin{eqnarray}}
\newcommand{\ear}{\end{eqnarray}}
\newcommand{\bt} { \begin{tabular} }
\newcommand{\et}{ \end{tabular} }
\newcommand{\bc} { \begin{center} }
\newcommand{\ec}{ \end{center} }

\title{Thermodynamics of antiferromagnetic alternating spin chains}
\author{G.A.P. Ribeiro\footnote{pavan@physik.uni-wuppertal.de} 
and A. Kl\"umper\footnote{kluemper@physik.uni-wuppertal.de} \\
Theoretische Physik, Bergische Universit\"at Wuppertal, \\ 42097 Wuppertal,
Germany}

\begin{document}

\maketitle
\thispagestyle{empty}
\begin{abstract}
  We consider integrable quantum spin chains with alternating spins
  $(S_1,S_2)$.  We derive a finite set of non-linear integral equations for
  the thermodynamics of these models by use of the quantum transfer matrix
  approach. Numerical solutions of the integral equations are provided for
  quantities like specific heat, magnetic susceptibility and in the case
  $S_1=S_2$ for the thermal Drude weight.  At low temperatures one class of
  models shows finite magnetization and the other class presents
  antiferromagnetic behaviour. The thermal Drude weight behaves linearly on $T$ at
  low temperatures and is proportional to the central charge $c$ of the
  system. Quite generally, we observe residual entropy for $S_1\neq S_2$. 
\end{abstract}

\centerline{PACS numbers: 05.50+q, 02.30.IK, 05.70Jk} \centerline{Keywords:
  Bethe Ansatz, Thermodynamics, Quantum transfer matrix, Mixed spin chain}

\newpage
\section{Introduction}
Integrable quantum systems and their associated classical vertex models have
been extensively studied in the last decades \cite{BAXTER,KOREPIN}. A large part of these systems is
exactly solvable by Bethe ansatz techniques providing spectral data and in
some cases also the eigenvectors.

After establishing the integrability and deriving the exact solution for the
spectrum, the main questions one likes to answer concern the physical properties of the system in
dependence on temperature, magnetic field etc.  There are many investigations
of integrable system in the thermodynamical limit at finite temperature. In
fact, we have several established routes to this goal. One may minimize the
free energy functional in the combinatorial Thermodynamical Bethe Ansatz
approach (TBA) \cite{YANG,TAKAHASHI,GAUDIN}, or one may apply algebraic and
analytical means for the computation of the partition function from the
quantum transfer matrix (QTM) \cite{MSUZUKI,KLUMPER92}.

The TBA approach is based on the string hypothesis and yields an infinite
set of non-linear integral equations (NLIE). However, it is 
impractical to solve the TBA equations numerically due to the infinite number
of equations and unknowns. Therefore approximations are required in this
approach.

By means of the quantum transfer matrix approach, a finite set of NLIE can be
derived exploiting analyticity properties of the quantum transfer matrix.
These equations have been shown to be successful in the description of
thermodynamical properties in the complete temperature range 
for many important models, like the Heisenberg model \cite{KLUMPER92,DEVEGA0,KLUMPER93}
and its spin-$S$ generalization \cite{JSUZUKI}, the $t-J$ model
\cite{KLUMPER-TJ}, the Hubbard model \cite{HUBBARD} and $SU(N)$ invariant
models for $N \leq 4$ \cite{KLUMPER-SUN}.

Nevertheless, the standard construction of the quantum transfer matrix assumes
models with isomorphic auxiliary and quantum spaces.  Here we are concerned
with extensions to more general models with non-isomorphic auxiliary and
quantum spaces. Important examples of such systems are mixed spin chains.
These mixed chains have been extensively studied for low and high temperatures
by use of the TBA equations and finite size scaling for isotropic chains
\cite{DEVEGA,ALADIM}. The dependence on magnetic fields was studied \cite{MARTINS,DEVEGA2,DOERFEL,FUJII} 
and more recently, also the anisotropic generalization was considered \cite{DOIKOU}.

Our aim is to propose a construction of the quantum transfer matrix by
replacing the standard ``rotation'' of vertex configurations of Boltzmann
weights by conjugated representations, i.e. by the normal Boltzmann weight
shifted by the crossing parameter. Having this in mind, we can tackle the more
general situation where the auxiliary and quantum spaces are not isomorphic.
As an application of this idea, we study the generic $(S_1,S_2)$ case of
alternating spin chains at finite temperature.

The paper is organized as follows. In section \ref{QTM}, we outline the basic
ingredients of the quantum transfer matrix approach. In section
\ref{alternating}, we define the alternating spin chain and its properties. In
section \ref{NLIE}, we derive the set of non-linear integral equations. In
section \ref{NUMERICAL}, we present our numerical findings for the solution of
the NLIE. Section \ref{thermal} is devoted to the calculation of the
thermal Drude weight for the case $S_1=S_2$. Our conclusions are given in
section \ref{CONCLUSION}.

\section{Quantum transfer matrix}
\label{QTM}

We are interested in the computation of the partition function 
$Z=\tr{e^{-\beta {\cal H}}}$ in the thermodynamical limit, on the condition
that $\cal H$ is an integrable local Hamiltonian derived from some row-to-row
transfer matrix.

In general, transfer matrices can be constructed as ordered products of many
different local Boltzmann weights 
${\cal L}_{{\cal A}i}(\lambda)$, where $\lambda$ denotes the spectral
parameter. These weights can be considered as matrices on the space $\cal A$,
usually called auxiliary space, which is related to the degrees of freedom on
the horizontal lines of a two dimensional vertex model. The matrix elements of
${\cal L}_{{\cal A}i}(\lambda)$ are operators acting non-trivially on the
site $i$ of the quantum space $\prod_{i=1}^{L} V_{i}$ of a chain of length $L$
and are related to the degrees of freedom on vertical lines.

The  product of Boltzmann weights
\eq {\cal T}_{\cal
  A}(\lambda)= {\cal L}_{{\cal A} L}(\lambda){\cal L}_{{\cal A}
  L-1}(\lambda)\dots {\cal L}_{{\cal A} 1}(\lambda),
\label{monodromy}
\en 
defines the monodromy matrix ${\cal T}_{\cal A}(\lambda)$. Here we allowed for
non-isomorphic spaces $V_i$. This way, ${\cal L}_{{\cal A} i}(\lambda)$ --
also called $\cal L$-operators-- may have 
different representations for the $L$ many quantum spaces ${\cal L}_{{\cal 
A} i}(\lambda)={\cal L}_{{\cal A} i}^{(\alpha,\beta_i)}(\lambda)$. The
labels for different representations, $\alpha,\beta_i$, may take for instance
integer values $\alpha,\beta_i=0,\dots,L-1$ and ${\cal L}_{{\cal A}
  i}^{(\alpha,\alpha)}(\lambda)$ denotes the isomorphic representation. Then
the row-to-row transfer matrix is the trace over the auxiliary space of the
monodromy matrix,
\eq
T(\lambda)=\tr_{\cal A}\left[ {\cal T}_{\cal A}(\lambda) \right].
\label{transfermatrix}
\en
The transfer matrix constitutes a family of commuting operators $\left[
  T(\lambda), T(\mu) \right]=0$, provided there is an invertible
$R$-matrix acting on the tensor product ${\cal A} \otimes {\cal A}$, such that
\eq 
R^{(\alpha)}(\lambda- \mu) {\cal L}_{{\cal A}i}^{(\alpha,\beta_{i})}(\lambda)
\otimes {\cal L}_{{\cal A}i}^{(\alpha,\beta_{i})}(\mu) = {\cal L}_{{\cal
    A}i}^{(\alpha,\beta_{i})}(\mu) \otimes {\cal L}_{{\cal
    A}i}^{(\alpha,\beta_{i})}(\lambda) R^{(\alpha)}(\lambda- \mu).
\label{fundrel}
\en
In order to have an associative algebra, the $R$-matrix is required to satisfy
the Yang-Baxter equation
\eq
R^{(\alpha)}_{12}(\lambda)R^{(\alpha)}_{23}(\lambda+\mu)R^{(\alpha)}_{12}(\mu)=
R^{(\alpha)}_{23}(\mu)R^{(\alpha)}_{12}(\lambda+\mu)R^{(\alpha)}_{23}(\lambda).
\label{yangbaxter}
\en

The simplest solution of (\ref{fundrel}) occurs when auxiliary and quantum
spaces $V_{i}$ are isomorphic implying that ${\cal
  L}_{12}^{(\alpha,\alpha)}(\lambda) = P_{12} R^{(\alpha)}_{12}(\lambda) $,
where $P_{12}$ is the permutation operator.

The conserved charges are obtained through the derivatives of the logarithm of
the transfer matrix
\eq 
{\cal J}^{(n)}=\frac{\partial^n}{\partial \lambda^n} \ln{\left[T(\lambda)
  \right]}\Big|_{\lambda=0},
\label{conserved}
\en
and the Hamiltonian corresponds to the first derivative, ${\cal H}={\cal
  J}^{(1)}$. Therefore, we can relate the transfer matrix and the Hamiltonian
in the following way
\eq
T(\lambda)=T(0) e^{\lambda {\cal H} + O(\lambda^2)},
\label{TH}
\en
where $T(0)$ plays the role of a kind of right multiple-step shift operator
\cite{DEVEGA} for a general distribution of $\cal L$-operators ${\cal
  L}_{{\cal A}i}^{(\alpha,\beta_{i})}(\lambda)$.

Let us consider that in addition to relation (\ref{fundrel}) the $\cal
L$-operators satisfy the following symmetry properties
\begin{align}
  \mbox{Unitarity: } & {\cal L}_{12}^{(\alpha,\beta)}(\lambda) {\cal
    L}_{12}^{(\alpha,\beta)}(-\lambda)= \zeta_{\alpha,\beta}(\lambda)
  \mbox{Id}_1  \otimes \mbox{Id}_2 , \label{uni} \\
  \mbox{Time reversal: } & {\cal
    L}_{12}^{(\alpha,\beta)}(\lambda)^{t_{1}}=
  {\cal L}_{12}^{(\alpha,\beta)}(\lambda)^{t_2},\label{tempo} \\
  \mbox{Crossing: } & {\cal L}_{12}^{(\alpha,\beta)}(\lambda) =
  \varsigma_{\alpha,\beta}(\lambda)M_1 {\cal
    L}_{12}^{(\alpha,\beta)}(-\lambda-\rho)^{t_{2}} M_{1}^{-1},
\label{cross}
\end{align}
where $\zeta_{\alpha,\beta}(\lambda)$ and $\varsigma_{\alpha,\beta}(\lambda)$
are scalar functions and $\rho$ is the crossing parameter. Here $\mbox{Id}_i$
and $t_{i}$ denote the identity matrix and transposition on the $i$-th space, 
$M_1=M \otimes \mbox{Id}_2$ where $M$ is some scalar matrix.

Now, we can define an adjoint transfer matrix $\overline{T}(\lambda)$ as
follows
\eq
\overline{T}(\lambda)= \prod_{i=1}^{L}\varsigma_{\alpha,\beta_i}(\lambda)
\tr_{\cal A}{\left[ {\cal L}_{{\cal A}
      L}^{(\alpha,\beta_{L})}(-\lambda-\rho){\cal L}_{{\cal A}
      L-1}^{(\alpha,\beta_{L-1})}(-\lambda-\rho)\dots {\cal L}_{{\cal A}
      1}^{(\alpha,\beta_{1})}(-\lambda-\rho) \right]},
\label{TMtilde}
\en
and by using the properties (\ref{tempo}-\ref{cross}) we can rewrite the
transfer matrix $\overline{T}(\lambda)$ such that,
\eq
\overline{T}(\lambda)=\tr_{\cal A}{\left[ {\cal L}_{{\cal A}
      1}^{(\alpha,\beta_{1})}(\lambda)\dots {\cal L}_{{\cal A}
      L-1}^{(\alpha,\beta_{L-1})}(\lambda) {\cal L}_{{\cal A}
      L}^{(\alpha,\beta_{L})}(\lambda) \right]}.
\en
Here we can see that, due to unitarity (\ref{uni}), the logarithmic derivative
results in the same Hamiltonian $\overline{\cal H}={\cal H}$ and
$\overline{T}(0)$ corresponds to the left multiple-step shift operator, such
that $T(0)\overline{T}(0)={\cal N}\mbox{ Id}$ where ${\cal N}=
\prod_{i=1}^{L}\zeta_{\alpha,\beta_i}(0)$.

In analogy to (\ref{TH}), we can write the transfer matrix
$\overline{T}(\lambda)$ as
\eq
\overline{T}(\lambda)=\overline{T}(0) e^{\lambda {\cal H} + O(\lambda^2)}.
\label{THbar}
\en

Using (\ref{TH}) and (\ref{THbar}) we can rewrite the partition function $Z$
in terms of the transfer matrices $T(\lambda)$  and  $\overline{T}(\lambda)$
by considering the Trotter limit,
\bear
Z&=&\lim_{N\rightarrow \infty} \tr{\left[ (e^{-\frac{2\beta}{N} {\cal H}}
    )^{N/2}\right]}, \\
 &=&\lim_{N\rightarrow \infty} \tr{\left[ \left(T(-\tau)\overline{T}(-\tau)
     \right)^{N/2}\right]}\frac{1}{{\cal N}^{N/2}},
\qquad\tau:=\frac{\beta}{N}.
\label{Ztrotter}
\ear

The partition function (\ref{Ztrotter}) can be related to a staggered vertex
model with alternating rows $T$ and $\overline{T}$. In this case we need to
know all the eigenvalues of these two transfer matrices to obtain the
partition function in a closed form. This is due to the fact that the
eigenvalues of both transfer matrices depend on the length of the quantum
chain $L$ and in particular on the Trotter number $N$, such that for
$N\to\infty$ all gaps close. However, we can circumvent this
problem by rewriting (\ref{Ztrotter}) in terms of the column-to-column
transfer matrix describing transfer in chain direction and hence is 
called the quantum transfer matrix 
\bear
\frac{T^{QTM}_{i}(x)}{(\varsigma_{\alpha,\beta_i}(-(\im x+\tau)))^{N/2}}&=&
\tr_{V_{i}}{[ {\cal L}_{V_{i} N}^{(\beta_i,\alpha)}(\im x+\tau-\rho){\cal
    L}_{V_{i}N-1}^{(\beta_i,\alpha)}(\im x-\tau) } \nonumber \\
&& \dots  {\cal L}_{V_{i}2}^{(\beta_i,\alpha)}(\im x+\tau-\rho) {\cal
  L}_{V_{i}1}^{(\beta_i,\alpha)}(\im x-\tau) ].
\label{QTMdef}
\ear

Each of these objects has a well defined largest eigenvalue separated by a gap
from the rest of the spectrum, even in the limit $N\to\infty$. Therefore, only
the largest eigenvalue is required for the computation of the partition
function. Here $x$ is the spectral parameter associated with the vertical line
ensuring the existence of a commuting family of matrices,
$\left[T^{QTM}_{i}(x),T^{QTM}_{i}(x')\right]=0$. However, of direct
physical relevance is $x=0$ for obtaining the partition function,
\eq
Z=\lim_{N\rightarrow \infty} \tr{\left[ \prod_{i=1}^{L}
    T_{i}^{QTM}(0) \right]}\frac{1}{{\cal N}^{N/2}}.
\label{Ztrotter2}
\en

Next, we address the identification of the largest eigenvalue of the product
of the quantum transfer matrices $T^{QTM}_{i}(x)$. In general, the
determination of the largest eigenvalue of the product of matrices
$\prod_{i=1}^{L}T^{QTM}_{i}(x)$ would require the knowledge of all the
eigenvalues of all transfer matrices $T^{QTM}_{i}(x)$, which could turn out to
be a more involved problem than the staggered model mentioned above.

Nevertheless, this problem can be overcome under certain conditions. For
instance, for the case of mixed spin chains all of the transfer matrices
commute according to the Yang-Baxter equation and the largest eigenvalues of
the individual transfer matrices correspond to the same eigenvector. This
implies that the largest eigenvalue of the product of $L$ different transfer
matrices is nothing than the product of the largest eigenvalues of the quantum
transfer matrices. In this work, we will restrict to this specific case.

Here we are interested in the free energy and its derivatives, so we have to
consider the logarithm of the partition function in the infinite length limit.
As the eigenvalues $\Lambda^{QTM}_{i}(x)$ depend only on the Trotter number,
we can first take the infinite length limit and later the infinite Trotter
number limit, which reads
\bear
f&=&-\frac{1}{\beta} \lim_{L,N\rightarrow \infty} \frac{1}{L}\ln{\left[ Z
  \right]}, \\
 &=&-\frac{1}{\beta} \lim_{N,L\rightarrow \infty} \frac{1}{L}
 \sum_{i=1}^{L} \ln{\left[ \Lambda_{i,max}^{QTM}(0)
   \right]}+\frac{1}{\beta}\lim_{N,L\rightarrow \infty}\frac{1}{L}
 \ln{\left[{\cal N}^{N/2} \right]}.
\label{freeEgeral}
\ear
Before closing this section, we would like to mention that the properties
(\ref{uni}-\ref{cross}) are also satisfied by many isomorphic self-crossed
models \cite{RESHETIKHIN1}. For the $SU(N)$ case with $N>2$, the property (\ref{cross})
reduces to the standard ``rotation" of the vertex configuration of the Boltzmann weights.

\section{Alternating spin chains}
\label{alternating}

In the previous section, we used unitarity, time reversal and crossing 
properties to construct the quantum transfer matrix considering general
representations of ${\cal L}_{{\cal A}i}^{(\alpha,\beta_i)}(\lambda)$. From
now on, we consider (for an even number of lattice sites $L$) the alternation
of two different representations of the group $SU(2)$ with spin $S_1$ at odd
sites and spin $S_2$ at even sites, i.e. $\beta_{2i-1}=S_1$ and
$\beta_{2i}=S_2$. In order to have a Hamiltonian with local interactions we
fix $\alpha$ to be identical to the spin $S_1$ representation (equivalently we
could have chosen $S_2$).

The monodromy matrix (\ref{monodromy}) becomes
\eq
{\cal T}_{\cal A}^{(S_1,S_2)}(\lambda)= {\cal L}_{{\cal A}
  L}^{(S_1,S_2)}(\lambda){\cal L}_{{\cal A} L-1}^{(S_1,S_1)}(\lambda)\dots
{\cal L}_{{\cal A} 2}^{(S_1,S_2)}(\lambda) {\cal L}_{{\cal A}
  1}^{(S_1,S_1)}(\lambda),
\label{mono-alter}
\en
with the auxiliary space ${\cal A}\equiv \C^{2S_1+1}$, and 
${\cal L}_{{\cal A} i}^{(S_1,S_2)}(\lambda)$ resp. 
${\cal L}_{{\cal A} i}^{(S_1,S_1)}(\lambda)$ are the $\cal L$-operators with spin
$S_1$ representation in the auxiliary space and $S_2$ resp. $S_1$ in the
quantum space.

The above $SU(2)$ invariant $\cal L$-operators can be obtained through the
fusion process \cite{KULISH}. Its explicit form conveniently normalized is
given by 
\eq
{\cal L}_{12}^{(S_1,S_2)}(\lambda)= \sum_{l=|S_1-S_2|}^{S_1+S_2}
f_{l}(\lambda) \check{P}_{l},
\label{Loperator}
\en
where\footnote{The symbol $*$ shall remind that the possibility $j=S_1-S_2$ is
  excluded throughout this work.}  $f_{l}(\lambda)=\prod_{j=l+1}^{S_1+S_2}
\left(\frac{\lambda- j}{\lambda+  j}\right)
{\prod^{*}}_{j=1}^{2S_{1}}(\lambda+S_2 -S_1 +j)$ and $\check{P}_l$ is the
projector onto the $SU(2)_l$ in the Clebsch-Gordon decomposition
$SU(2)_{S_1}\otimes SU(2)_{S_2}$. This operator is represented by
\eq
\check{P}_{l}=\prod_{\stackrel{k=|S_1-S_2|}{k \neq l}}^{S_1+S_2}
\frac{\vec{S}_{1}\otimes \vec{S}_{2}-x_{k}}{x_{l}-x_{k}},
\en
with $x_{l}=\frac{1}{2}\left[l(l+1)-S_1(S_1+1)-S_2(S_2+1)\right]$ and the
$SU(2)$ generators
$\vec{S}_{a}=(\hat{S}_{a}^{x},\hat{S}_{a}^{y},\hat{S}_{a}^{z})$ for $a=1,2$.

The operator (\ref{Loperator}) is a solution of (\ref{fundrel}) with the
following $R$-matrix 
\eq
R_{12}^{(S_1)}(\lambda)=P_{12}{\cal L}_{12}^{(S_1,S_1)}(\lambda).
\en
It satisfies the properties (\ref{uni}-\ref{cross}) with scalar functions
given by $\zeta_{S_1,S_2}(\lambda)=\prod_{j=1}^{2S_1} ((S_2-S_1+j)^2 -
\lambda^2)$ and $\varsigma_{S_1,S_2}(\lambda)= (-1)^{2S_1}$ and crossing
parameter $\rho=1$. The matrix $M$ is an anti-diagonal matrix whose non-zero
elements are $M_{i,j}=-(-1)^{i}\delta_{i,2S_{1}+2-j}$.

The Hamiltonian associated to the transfer matrix $T(\lambda)=\tr_{\cal
  A}{\left[ {\cal T}_{\cal A}^{(S_1,S_2)}(\lambda) \right]}$ has terms with
two and three site interactions. Its generic expression is given by
\bear
{\cal H}^{(S_1,S_2)}&=&\sum_{\text{even } i} \left[{\cal
    L}_{i-1,i}^{(S_1,S_2)}(0)\right]^{-1} \frac{\partial}{\partial\lambda}{\cal L}_{i-1,i}^{(S_1,S_2)}(\lambda) \Big|_{\lambda=0}
 \\
&+&\sum_{\text{odd } i}  \left[{\cal L}_{i-2,i-1}^{(S_1,S_2)}(0)\right]^{-1}
\left[{\cal L}_{i-2,i}^{(S_1,S_1)}(0)\right]^{-1} \frac{\partial}{\partial\lambda} {\cal
  L}_{i-2,i}^{(S_1,S_1)}(\lambda)\Big|_{\lambda=0} {\cal L}_{i-2,i-1}^{(S_1,S_2)}(0),\nonumber
\ear
where periodic boundary conditions are assumed. For illustration, the
Hamiltonian for case $S_1=1/2$, $S_2=S$ is given explicitly by \cite{ALADIM}
\bear
{\cal H}^{(\frac{1}{2},S)}&=&\frac{1}{2}\left(\frac{1}{S+\frac{1}{2}}
\right)^{2} \Big[\sum_{\text{even }i}\left( \vec{\sigma}_{i-1}\cdot
  \vec{S}_{i} + \vec{S}_{i} \cdot \vec{\sigma}_{i+1} +
  \left\{\vec{\sigma}_{i-1}\cdot \vec{S}_{i} ,\vec{S}_{i} \cdot
    \vec{\sigma}_{i+1}   \right\} \right)\nonumber \\ 
&+&\left( \frac{1}{4}-S(S+1) \right)\sum_{\text{even }i}  \vec{\sigma}_{i-1}
\cdot \vec{\sigma}_{i+1}\Big]  + \frac{L}{4}\left(
  1+\frac{1}{(S+\frac{1}{2})^2}\right).
\label{hamilt1o2S}
\ear

One of the consequences of the alternation of two different spins is that we
have two quantum transfer matrices to work with. We denote them by
$T^{(S_1)}(x)$ and $T^{(S_2)}(x)$, such as
\bear
T^{(S_a)}(x)&:=&T^{QTM}_a(x)=\tr_{V_{a}}{[ {\cal L}_{V_{a} N}^{(S_a,S_1)}(\im
  x+\tau-\rho){\cal L}_{V_{a}N-1}^{(S_a,S_1)}(\im x-\tau) } \nonumber \\
&&\dots {\cal L}_{V_{a}2}^{(S_a,S_1)}(\im x+\tau-\rho) {\cal
  L}_{V_{a}1}^{(S_a,S_1)}(\im x-\tau) ],
\label{QTMalter}
\ear
where the vertical spaces are $V_a \equiv \C^{2S_a+1}$ and $a=1,2$.

The transfer matrices (\ref{QTMalter}) for $a=1,2$ commute due to the
Yang-Baxter relation \cite{BABUJIAN}. Therefore, they can be diagonalized
simultaneously. It can also be deduced from \cite{BABUJIAN} that their largest
eigenvalues correspond to the same eigenstate. Hence the largest eigenvalue of
the product $T^{(S_1)}(x)T^{(S_2)}(x)$ is the product of the largest
eigenvalues of $T^{(S_1)}(x)$ and $T^{(S_2)}(x)$.

For the analysis of the spectra we use the fusion hierarchy for the quantum
transfer matrix $T^{(j)}(x)$, in analogy to the fusion of $\cal
L$-operators. The algebraic relations read (see e.g. \cite{JSUZUKI})
\begin{align}
&T^{(j)}(x)T^{(\frac{1}{2})}(x+\im(j+\frac{1}{2}))= a_j(x)
T^{(j+\frac{1}{2})}(x+\frac{\im}{2}) +
a_{j+1}(x)T^{(j-\frac{1}{2})}(x-\frac{\im}{2}), \nonumber\\
&T^{(0)}(x)=a_0(x) \mbox{Id}, ~~ j=\frac{1}{2},1,\frac{3}{2},\dots 
\label{fusionh}
\end{align}
where $a_j(x)=\prod_{l=1}^{2
  S_1}\phi_{+}(x+\im(j-S_1+l-1))\phi_{-}(x+\im(j-S_1+l))$ and
$\phi_{\pm}(x)=(x \pm \im \tau)^{N/2}$.

From the fusion hierarchy with bilinear and linear expressions in $T$
(\ref{fusionh}), one can obtain another set of functional relations
\cite{KPEARCE}, usually called $T$-system, with exclusively bilinear
expressions
\bear
T^{(j)}(x+\frac{\im}{2})T^{(j)}(x-\frac{\im}{2})=
T^{(j-\frac{1}{2})}(x) T^{(j+\frac{1}{2})}(x) +f_{j}(x) \mbox{
  Id},
\label{Tsystem}
\ear
where $f_{j}(x)=\prod_{l=1}^{2
  S_1}\phi_{+}(x-\im(j-S_{1}+l+\frac{1}{2}))\phi_{-}(x-\im(j-S_{1}+l-
\frac{1}{2}))
\phi_{+}(x+\im(j-S_{1}+l-\frac{1}{2}))\phi_{-}(x+\im(j-S_{1}+l+\frac{1}{2})) $ 
for any $j$ integer or semi-integer.

Equally important is a set of functional relations referred to as the
$Y$-system, which is a consequence of (\ref{Tsystem}). It is written as
\eq
y^{(j)}(x+\frac{\im}{2})y^{(j)}(x-\frac{\im}{2})=
Y^{(j-\frac{1}{2})}(x)Y^{(j+\frac{1}{2})}(x),
\label{Ysystem}
\en
where $y^{(j)}(x) =
\frac{T^{(j-\frac{1}{2})}(x)T^{(j+\frac{1}{2})}(x)}{f_j(x)}$ and
$Y^{(j)}(x)=1+y^{(j)}(x)$.

Lastly, we introduce a Zeeman term $\widetilde{\cal H}={\cal H}- h
\hat{S}^z$. This term represents the coupling of the magnetic field $h$ to
the spin $\hat{S}^z=\sum_{\stackrel{i=1}{\text{odd }i}}^L \hat{S}_{1,i}^z
+\sum_{\stackrel{i=1}{\text{even } i}}^L \hat{S}_{2,i}^{z}$. It can be
introduced inside the trace of the partition function such as,
\eq
Z=\lim_{N\rightarrow \infty} \tr{\left[ \left(T(-\tau)\overline{T}(-\tau)
    \right)^{N/2} e^{\beta h \hat{S}^z}\right]}\frac{1}{{\cal N}^{N/2}}.
\en
Alternatively, it can be considered as a diagonal boundary term on the
vertical lines along a horizontal seam. This redefines only trivially the
quantum transfer matrix 
\bear
T^{(S_a)}(x)&=&\tr_{V_{a}}{[ {\cal G}_a {\cal L}_{V_{a} N}^{(S_a,S_1)}(\im
  x+\tau-\rho){\cal L}_{V_{a}N-1}^{(S_a,S_1)}(\im x-\tau) } \nonumber \\
&&\dots {\cal L}_{V_{a}2}^{(S_a,S_1)}(\im x+\tau-\rho) {\cal
  L}_{V_{a}1}^{(S_a,S_1)}(\im x-\tau) ],
\label{QTMalterG}
\ear
where ${\cal G}_a$ is a diagonal matrix whose non-zero elements are $({\cal
  G}_a)_{i,i}=e^{\beta h (S_a+1-i)}$.

The eigenvalues $\Lambda^{(j)}(x)$ associated to $T^{(j)}(x)$ also
satisfy the functional relations (\ref{fusionh}-\ref{Ysystem}). This is due to
the commutativity property among different $T^{(j)}(x)$. This way, we
obtain the eigenvalues at any fusion level in terms of the first level
eigenvalue through the iteration of the relations (\ref{fusionh}) and
(\ref{Tsystem}). Alternatively, we can proceed along the same lines as
\cite{MELO} applying the algebraic Bethe ansatz to the case of twisted
boundary conditions.

In both cases we end up with the eigenvalues of the
quantum transfer matrix (\ref{QTMalterG}), 
\bear
\Lambda^{(j)}(x)=\sum_{m=1}^{2j+1} \lambda_{m}^{(j,S_1)}(x), 
\ear 
\eq
\lambda_{m}^{(j)}(x)=e^{\beta h(j+1-m)} t_{+,m}^{(j)}(x)
t_{-,m}^{(j)}(x+\im)
\frac{Q(x-\im(\frac{1}{2}+j))Q(x+\im(\frac{1}{2}+j))}
{Q(x-\im(\frac{3}{2}+j-m))Q(x-\im(\frac{1}{2}+j-m))},
\label{Elambda}
\en
where $\displaystyle t_{\pm,m}^{(j)}(x)=\prod_{l=j-m+2}^{j}
\frac{\phi_{\pm}(x-\im(l-S_1))}{\phi_{\pm}(x-\im(l+S_1))}
{\prod_{l=1}^{2S_{1}}}^{*}\phi_{\pm}(x-\im(j -S_1 +l))$ and
$Q(x)=\prod_{l=1}^{n}(x-x_{l})$. The corresponding Bethe ansatz equations can
be written as 
\eq
e^{\beta h}\frac{\phi_{+}(x_l-\im(S_1+\frac{1}{2}))
  \phi_{-}(x_l-\im(S_1-\frac{1}{2}))}{\phi_{-}(x_l+\im(S_1+\frac{1}{2}))
  \phi_{+}(x_l+\im(S_1-\frac{1}{2}))}=\prod_{\stackrel{j=1}{j\neq l}}^{n}
\frac{x_l-x_j-\im}{x_l-x_j+\im}.
\label{BAeq}
\en

According to the previous section, we only need to know the largest eigenvalue
in the limit $N\rightarrow \infty$ to describe the thermodynamics of the one
dimensional quantum model. Then for instance by numerical analysis of the
Bethe ansatz equation (\ref{BAeq}) for small $N$ we see that the largest
eigenvalue lies in the sector $n=S_1 N$. However, the limit $N\rightarrow
\infty$ cannot be considered numerically. So, we need to encode the Bethe
ansatz roots in such a way that the free energy can be evaluated independently
of the exact knowledge of the individual roots.

One possible way is to define a set of suitable auxiliary functions depending on the Bethe ansatz roots. Then by
exploiting the above and further functional relations we eliminate the
explicit dependence on the roots. Therefore the Bethe ansatz roots for finite
$N$ (including the limit $N\rightarrow \infty$) become encoded in a finite set
of auxiliary functions satisfying certain non-linear integral equations.

Such an analysis was already done for many cases, for instance for the
spin-$1/2$ Heisenberg chain \cite{KLUMPER92,DEVEGA0,KLUMPER93} and its higher spin
extensions \cite{JSUZUKI}. In the latter case, the auxiliary functions were
taken as a subset of the $y$-functions complemented by two ``novel'' functions
which reduce the infinitely many functional relations (\ref{Ysystem}) to 
finitely many. This is the starting point of the next section.

\section{Non-linear integral equations}
\label{NLIE}

In this section, we introduce a suitable set of auxiliary functions and
explore its analyticity properties to obtain a finite set of non-linear
integral equations. These auxiliary functions turn out to describe the largest
eigenvalue of (\ref{QTMalterG}) and consequently the free energy
(\ref{freeEgeral}) at finite temperature.
Specifically, we need to define $2 s + 1$ auxiliary functions, where
$s=\text{max}(S_1,S_2)$. We will proceed along the lines of \cite{JSUZUKI} and
take as the first $2s-1$ auxiliary functions the $y$-functions
\eq
y^{(j)}(x) =
\frac{\Lambda^{(j-\frac{1}{2})}(x)\Lambda^{(j+\frac{1}{2})}(x)}{f_j(x)},
~~ j=\frac{1}{2},\dots,s-\frac{1}{2}.
\label{auxy}
\en
The two remaining functions are defined as
\bear 
b(x)&=&\frac{\lambda_1^{(s)}(x+\frac{\im}{2}) + \dots + \lambda_{2
    s}^{(s)}(x+\frac{\im}{2})
}{\lambda_{2s+1}^{(s)}(x+\frac{\im}{2})}, \label{auxb}\\ 
\bar{b}(x)&=&\frac{\lambda_2^{(s)}(x-\frac{\im}{2}) + \dots + \lambda_{2
    s+1}^{(s)}(x-\frac{\im}{2})
}{\lambda_{1}^{(s)}(x-\frac{\im}{2})}. \label{auxbbar}
\ear 
In addition to this, we introduce a shorthand notation for simply related
functions $B(x):=1+b(x)$, $\bar{B}(x):=1+\bar{b}(x)$ and
$Y^{(j)}(x):=1+y^{(j)}(x)$ for $j=\frac{1}{2},\dots,s-\frac{1}{2}$.

In conformity with the previous definition, we note that $B(x)=
\frac{\Lambda^{(s)}(x+\frac{\im}{2})}{\lambda_{2
    s+1}^{(s)}(x+\frac{\im}{2})}$ and
$\bar{B}(x)=\frac{\Lambda^{(s)}(x-\frac{\im}{2})}{\lambda_1^{(s)}
(x-\frac{\im}{2})
}$ with product $B(x)\bar{B}(x) = Y^{(s)}(x)$. This implies for
the first $(2 s-1)$ functional relations (\ref{Ysystem}) 
\begin{align}
&y^{(j)}(x+\frac{\im}{2})y^{(j)}(x-\frac{\im}{2})=
Y^{(j-\frac{1}{2})}(x)Y^{(j+\frac{1}{2})}(x)
~ \mbox{for }
j=\frac{1}{2},1,\dots,s-1, \label{yrelat}\\
&y^{(s-\frac{1}{2})}(x+\frac{\im}{2})y^{(s-\frac{1}{2})}
(x-\frac{\im}{2})=Y^{(s-1)}(x)B(x)\bar{B}(x).  
\label{yBrelat}
\end{align}

We can write $b(x), \bar{b}(x), B(x)$ and $\bar{B}(x)$ explicitly using
(\ref{Elambda}) such that
\begin{align}
&b(x)=\frac{Q(x+\im(s+1))}{Q(x-\im s)}\frac{e^{\beta h
    (s+\frac{1}{2})}\Lambda^{(s-\frac{1}{2})}(x)}{\prod_{l=1}^{2 S_1}
  \phi_{+}(x+\im(s-S_1+l-\frac{1}{2}))\phi_{-}(x+\im(s-S_1+l+\frac{1}{2}))},
\label{baux} \\
&\bar{b}(x)=\frac{Q(x-\im(s+1))}{Q(x+\im s)}\frac{e^{-\beta h
    (s+\frac{1}{2})}\Lambda^{(s-\frac{1}{2})}(x)}{\prod_{l=1}^{2 S_1}
  \phi_{+}(x-\im(s-S_1+l+\frac{1}{2}))\phi_{-}(x-\im(s-S_1+l-\frac{1}{2}))},
\label{bbaraux}\\
&B(x)= \frac{Q(x+\im s)}{Q(x-\im s)} \frac{e^{\beta h
    s}\Lambda^{(s)}(x+\frac{\im}{2})}{\prod_{l=1}^{2 S_1}
  \phi_{+}(x+\im(s-S_1+l-\frac{1}{2}))\phi_{-}(x+\im(s-S_1+l+\frac{1}{2}))},
\label{auxcb1} \\
&\bar{B}(x)=\frac{Q(x-\im s)}{Q(x+\im s)}\frac{e^{-\beta h
    s}\Lambda^{(s)}(x-\frac{\im}{2})}{\prod_{l=1}^{2 S_1}
  \phi_{+}(x-\im(s-S_1+l+\frac{1}{2}))\phi_{-}(x-\im(s-S_1+l-\frac{1}{2}))}. 
\label{auxcb}
\end{align}
In this way, it is evident that $b(x)$, $\bar{b}(x)$ are related to
$\Lambda^{(s-\frac{1}{2})}(x)$.

Moreover,
$\Lambda^{(s-\frac{1}{2})}(x)$ is related to $Y^{(s-\frac{1}{2})}(x)$
through the definition of $y$-function. This relation can be written as
\eq
\Lambda^{(s-\frac{1}{2})}(x+\frac{\im}{2})\Lambda^{(s-\frac{1}{2})}
(x-\frac{\im}{2})=f_{s-\frac{1}{2}}(x)Y^{(s-\frac{1}{2})}(x). 
\label{auxY}
\en
At this point, we have a common set of functions which still depend on the
Bethe ansatz roots and whose limit $N\rightarrow \infty$ is still to be
performed. However, this dependence as well as the limit can be worked out
easily in Fourier space.

In order to calculate the Fourier transform, we exploit the analyticity
properties of the eigenvalue of the quantum transfer matrix and the auxiliary
functions.  Furthermore, these functions should be non-zero and have constant
asymptotics in a strip around the real axis. This allows us to apply the
Fourier transform to the logarithmic derivative of the auxiliary functions,
\eq 
\hat{f}(k)=\int_{-\infty}^{\infty} \frac{d}{d x}\left[ \llg{f(x)} \right]
e^{-\im k x} \frac{dx}{2\pi}.
\label{FT}
\en 
In the cases $k<0$ and $k>0$, we have chosen a closed contour above and below
the real axis, respectively. For this reason, it is of fundamental importance
to analyze the structure of the zeros of the auxiliary functions.

In particular, the zeros and poles of the auxiliary functions
(\ref{auxy}-\ref{auxbbar}) originate from the zeros of $Q(x)$ and
$\Lambda^{(j)}(x)$ for $j=\frac{1}{2},\dots,s$ besides those of the
$\phi_{\pm}(x)$ functions. Therefore, we have to analyze the qualitative
distribution of the Bethe ansatz roots as well as the zeros of the eigenvalue functions $\Lambda^{(j)}(x)$.

It is well known that Bethe ansatz roots form $2 S_1$-strings in the particle
sector $n=S_1 N$. These roots have imaginary parts placed approximately at $(S_1+\frac{1}{2}-l)$
for $l=1,\dots,2 S_1$ \cite{BABUJIAN}. Concerning the zeros of
$\Lambda^{(j)}(x)$ for $j=\frac{1}{2},\dots,s$, we have verified numerically
that their imaginary parts are placed at $\pm(j-S_1+l)$ for $l=1,\dots,2 S_1$ and $l\neq S_1-j$.

By direct inspection of (\ref{auxy},\ref{yrelat}-\ref{auxY}), we note that
almost all auxiliary functions are free of zeros and poles in a strip
containing $-1/2 \leq \Im(x) \leq 1/2$. The exceptions are $y^{(S_1)}(x)$ for
$S_1 <S_2$ and $b(x),\bar{b}(x)$ for $S_1 \geq S_2$, which should be treated
separately.

This way, the position of the zeros and poles of the auxiliary functions
depend on the relative magnitude of $S_1$ and $S_2$. So, we have to split our
analysis in three parts: $S_1 < S_2$, $S_1 = S_2$ and $S_1 > S_2$.

\subsection{$S_1 < S_2$}

In this case, we have $s=S_2$ in the previous definition. In order to deal
with the problem involving the function $y^{(S_1)}(x)$, we define a related
function for which the problematic zeros and poles at $x=\pm \im/2$ are
cancelled,
\eq
\tilde{y}^{(S_1)}(x)=\frac{\phi_{+}(x+\frac{\im}{2})\phi_{-}
(x-\frac{\im}{2})}{\phi_{-}(x+\frac{\im}{2})\phi_{+}(x-\frac{\im}{2})} 
y^{(S_1)}(x).
\label{ytilde}
\en
Consequently, the $2 S_1$-th equation in (\ref{yrelat}) becomes
\eq
\tilde{y}^{(S_1)}(x+\frac{\im}{2})\tilde{y}^{(S_1)}(x-\frac{\im}{2})=
\frac{\phi_{-}(x-\im)\phi_{+}(x+\im)}{\phi_{+}(x-\im)\phi_{-}(x+\im)}
Y^{(S_1-\frac{1}{2})}(x)Y^{(S_1+\frac{1}{2})}(x),
\label{tyrelat}
\en
and the functions $\tilde{y}^{(S_1)}(x\pm\frac{\im}{2})$ can be
transformed as usual according to (\ref{FT}). On the other hand, we can apply
the Fourier transform to the equation (\ref{ytilde}), once it does not have
zeros and poles on the real axis. Thus we are able to establish a relation
between $y^{(S_1)}$ and $\tilde{y}^{(S_1)}$ in Fourier space,
\eq
\hat{\tilde{y}}^{(S_1)}(k)=\im N\sinh{\left[k\beta/N \right]} e^{-|k|/2} 
+ \hat{y}^{(S_1)}(k).
\label{ytyrel}
\en

Now, applying (\ref{FT}) to the functional relations (\ref{yrelat}-\ref{auxY}) and (\ref{tyrelat}) we obtain after a long but
straightforward calculation a set of algebraic relations in Fourier space.
These relation are given in terms of the transformed auxiliary functions
$\hat{y}^{(j)}(k)$, $\hat{b}(k)$, $\hat{\bar{b}}(k)$, $\hat{Y}^{(j)}(k)$,
$\hat{B}(k)$, $\hat{\bar{B}}(k)$ and the unknowns
$\hat{\Lambda}^{(S_2-\frac{1}{2})}(k)$, $\hat{\Lambda}^{(S_2)}(k)$ and
$\hat{Q}(k)$. We can eliminate the unknowns after some algebraic manipulation.
Finally, using (\ref{ytyrel}) we obtain
\eq
\left(
\begin{array}{c}
\hat{y}^{(\frac{1}{2})}(k) \\
\vdots \\
\hat{y}^{(S_1)}(k) \\
\vdots \\
\hat{y}^{(S_2-\frac{1}{2})}(k) \\
\hat{b}(k) \\
\hat{\bar{b}}(k)
\end{array}\right)=
\left(\begin{array}{c}
0 \\
\vdots \\
\hat{d}(k) \\
\vdots \\
0 \\
0 \\
0
\end{array}\right)
+
\hat{{\cal K}}(k)
\left(\begin{array}{c}
\hat{Y}^{(\frac{1}{2})}(k) \\
\vdots \\
\hat{Y}^{(S_1)}(k) \\
\vdots \\
\hat{Y}^{(S_2-\frac{1}{2})}(k) \\
\hat{B}(k) \\
\hat{\bar{B}}(k)
\end{array}\right),
\label{eqS2S1FS}
\en
where the kernel $\hat{\cal K}(k)$ is a $(2 S_2+1)\times (2 S_2+1) $ matrix
given by
\eq
\hat{\cal K}(k)=
\left(
\begin{array}{cccccccc}
0 & \hat{K}(k) & 0 & \cdots & 0 & 0 & 0 & 0 \\
\hat{K}(k) & 0 &  \hat{K}(k)&   & \vdots & \vdots & \vdots & \vdots  \\
0 & \hat{K}(k) & 0  &  &  & 0 & 0 & 0 \\
\vdots &  &   &   & 0 & \hat{K}(k) & 0 & 0 \\
0 & 0 & \cdots  & 0 & \hat{K}(k) & 0 & \hat{K}(k) & \hat{K}(k) \\
0 & 0 & \cdots  & 0 & 0 & \hat{K}(k) & \hat{F}(k) & -e^{-k}\hat{F}(k) \\
0 & 0 & \cdots  & 0 & 0 & \hat{K}(k) & -e^{k}\hat{F}(k) & \hat{F}(k) \\
\end{array}\right),
\label{Kernel-k}
\en
with $\hat{K}(k)=\frac{1}{2\cosh{\left[k/2\right]}}$,
$\hat{F}(k)=\frac{e^{-|k|/2}}{2\cosh{\left[k/2\right]}}$ and $\hat{d}(k)=-\im
N \frac{\sinh{\left[ k\beta/N\right]}}{2\cosh{\left[k/2\right]}}$.

As the Trotter number $N$ appears only in $\hat{d}(k)$, we can take the limit
$N\rightarrow\infty$ straightforwardly,
\eq
\hat{d}(k)=-\frac{\im}{2\cosh{\left[k/2 \right]}} \lim_{N\rightarrow \infty} N
\sinh{\left[ k\beta/N \right]}=- \frac{\im k\beta}{2\cosh{\left[ k/2\right]}}.
\en

The inverse Fourier transform has been applied to (\ref{eqS2S1FS}) followed by
an integration over $x$, resulting in
\eq
\left(
\begin{array}{c}
\llg{y^{(\frac{1}{2})}(x)} \\
\vdots \\
\llg{y^{(S_1)}(x)} \\
\vdots \\
\llg{y^{(S_2-\frac{1}{2})}(x)} \\
\llg{b(x)} \\
\llg{\bar{b}(x)}
\end{array}\right)=
\left(\begin{array}{c}
0 \\
\vdots \\
- \beta d(x) \\
\vdots \\
0 \\
\beta \frac{h}{2}  \\
-\beta \frac{h}{2}
\end{array}\right)
+
{\cal K}*
\left(\begin{array}{c}
\llg{Y^{(\frac{1}{2})}(x)} \\
\vdots \\
\llg{Y^{(S_1)}(x)} \\
\vdots \\
\llg{Y^{(S_2-\frac{1}{2})}(x)} \\
\llg{B(x)} \\
\llg{\bar{B}(x)}
\end{array}\right),
\en
where $d(x)=\frac{\pi}{\cosh{\left[ \pi x \right]}}$ and the symbol $*$
denotes the convolution $f*g(x)=\int_{-\infty}^{\infty} f(x-y)g(y)dy$. The
integration constants $\pm \beta h/2$ were determined in the asymptotic limit
$|x|\rightarrow \infty$.

The kernel matrix is given explicitly by
\eq
{\cal K}(x)= 
\left(
\begin{array}{cccccccc}
0 & K(x) & 0 & \cdots & 0 & 0 & 0 & 0 \\
K(x) & 0 &  K(x) &   & \vdots & \vdots & \vdots & \vdots  \\
 0 & K(x) & 0  &  &  & 0 & 0 & 0 \\
\vdots &  &   &   & 0 & K(x) & 0 & 0 \\
0 & 0 & \cdots  & 0 & K(x) & 0 & K(x) & K(x) \\
0 & 0 & \cdots  & 0 & 0 & K(x) & F(x) & -F(x+\im) \\
0 & 0 & \cdots  & 0 & 0 & K(x) & -F(x-\im) & F(x) \\
\end{array}\right),
\label{Kernel-x}
\en
where $K(x)=\frac{\pi}{\cosh{\left[ \pi x \right]}}$ and
$F(x)=\int_{-\infty}^{\infty}\frac{e^{-|k|/2+\im k x}}{2 \cosh{\left[ k/2
    \right]}} dk $.

Now, we have to derive an expression for the eigenvalue
$\Lambda^{(S_2)}(x)$ in terms of the auxiliary functions.
It is convenient to define a new function
\eq
\underline{\Lambda}^{(S_2)}(x)=\frac{\Lambda^{(S_2)}(x)}{\prod_{l=1}^{2
    S_1}\phi_{+}(x-\im(S_2-S_1+l)) \phi_{-}(x+\im(S_2-S_1+l))},
\label{eingnorm}
\en
which has constant asymptotics. For $x=0$ and finite $N$, we have
$\llg{\Lambda^{(S_2)}}(0)=\llg{\underline{\Lambda}^{(S_2)}}(0)+\sum_{l=1}^{2
  S_1}\llg{\left[1-\frac{\beta}{S_2-S_1+l}\frac{1}{N}\right]^{N}}+
\frac{2}{L}\llg{\left[{\cal N}^{N/2}\right]}$, where we have used the fact
that ${\cal N}=\prod_{l=1}^{2 S_1}(S_2-S_1+l)^L$.

Using the Fourier transformed version of
(\ref{auxcb1}-\ref{auxcb},\ref{eingnorm}), we obtain
\eq
\hat{\underline{\Lambda}}^{(S_2)}(k)= \im k\beta
\frac{e^{-|k|(S_2-S_1-\frac{1}{2})}}{2\cosh{\left[k/2\right]}} \sum_{l=1}^{2
  S_1}e^{-|k|l} + \hat{K}(k) \left[\hat{B}(k) + \hat{\bar{B}}(k) \right].
\en

Proceeding as before, we apply the inverse Fourier transform followed by an
integration over $x$ and the determination of the integration
constant. In this way, we obtain
\eq
\llg{\underline{\Lambda}^{(S_2)}(x)}=\beta \epsilon^{(S_2,S_1)}(x) +
\left(K*\llg{B\bar{B}}\right)(x),
\en
where $\epsilon^{(S_2,S_1)}(x)$ is given by
\eq
\epsilon^{(S_2,S_1)}(x)=\sum_{l=1}^{2 S_1} \int_{-\infty}^{\infty}
\frac{e^{-|k|(S_2-S_1+l-\frac{1}{2})}}{2\cosh{\left[k/2\right]}}  e^{\im k x}
dk .
\en

At the point $x=0$, we can rewrite this integral in terms of the Euler psi
function,
\eq
\epsilon^{(S_2,S_1)}(0)=\psi\left(\frac{S_2+S_1+1}{2}\right)-
\psi\left(\frac{S_2-S_1+1}{2}\right).
\label{S2S1GS}
\en

The contribution of the quantum transfer matrix $T^{(S_2)}(0)$
(\ref{QTMalterG}) to the free energy is given by (\ref{freeEgeral})
\bear
f^{(S_2,S_1)}&=&-\frac{1}{2\beta}\lim_{N \rightarrow \infty}
\llg{\Lambda^{(S_2)}(0)}+ \frac{1}{\beta}\lim_{N,L\rightarrow \infty} \frac{1}{L}\llg{\left[{\cal N}^{N/2}\right]},  \\
&=&-\frac{1}{2\beta}\lim_{N\rightarrow \infty}
\llg{\underline{\Lambda}^{(S_2)}(0)}+\frac{1}{2}\sum_{l=1}^{2
  S_1}\frac{1}{S_2-S_1+l}.
\ear
Therefore, we can write $f^{(S_2,S_1)}$ explicitly as
\bear
f^{(S_2,S_1)}&=&\frac{1}{2}\left[\sum_{l=1}^{2 S_1}\frac{1}{S_2-S_1+l}
  -\psi\left(\frac{S_2+S_1+1}{2}\right)+\psi\left(\frac{S_2-S_1+1}{2}\right)
\right]\nonumber\\
&&-\frac{1}{2\beta}\left(K*\llg{B\bar{B}}\right)(0).
\label{energyS2S1}
\ear

\subsection{$S_1 = S_2$}

In this case, we note that $b(x)$ and $\bar{b}(x)$ have zeros at $x=\pm \im/2$
which are presenting some subtleties. These zeros originate from the factor
$\Lambda^{(S_1-\frac{1}{2})}$ and in principle do not present any problems for
the computation of the Fourier transform of the logarithmic derivative of
(\ref{baux}-\ref{bbaraux}).  The problem arises in the Fourier transform
of (\ref{auxY}), which is required to eliminate the unknown function
$\Lambda^{(S_1-\frac{1}{2})}$.

Hence, we define a new function
$\widetilde{\Lambda}^{(S_1-\frac{1}{2})}(x)=\frac{\Lambda^{(S_1-
    \frac{1}{2})}(x)}{\phi_{+}(x-\im/2)\phi_{-}(x+\im/2)}$, which does not
have any zeros at $x=\pm \im/2$. We apply (\ref{FT}) to the functional
relations (\ref{yrelat}-\ref{auxY}) with $\widetilde{\Lambda}$ instead of
${\Lambda}$. Then we eliminate the unknowns
$\hat{\Lambda}^{(S_1-\frac{1}{2})}(k)$ and $\hat{Q}(k)$ and finally we
obtain
\eq
\left(
\begin{array}{c}
\hat{y}^{(\frac{1}{2})}(k) \\
\vdots \\
\hat{y}^{(S_1-\frac{1}{2})}(k) \\
\hat{b}(k) \\
\hat{\bar{b}}(k)
\end{array}\right)=
\left(\begin{array}{c}
0 \\
\vdots \\
0 \\
\hat{d}(k) \\
\hat{d}(k) 
\end{array}\right)
+
\hat{{\cal K}}(k)
\left(\begin{array}{c}
\hat{Y}^{(\frac{1}{2})}(k) \\
\vdots \\
\hat{Y}^{(S_1-\frac{1}{2})}(k) \\
\hat{B}(k) \\
\hat{\bar{B}}(k)
\end{array}\right),
\label{eqS1S1FS}
\en
where the kernel $\hat{\cal K}(k)$ with the same structure as (\ref{Kernel-k}), is a $(2 S_1+1)\times (2 S_1+1) $ matrix.

Applying the inverse Fourier transform to (\ref{eqS1S1FS}) followed by an
integration over $x$, results in
\eq
\left(
\begin{array}{c}
\llg{y^{(\frac{1}{2})}(x)} \\
\vdots \\
\llg{y^{(S_1-\frac{1}{2})}(x)} \\
\llg{b(x)} \\
\llg{\bar{b}(x)}
\end{array}\right)=
\left(\begin{array}{c}
0 \\
\vdots \\
0 \\
- \beta d(x) +\beta \frac{h}{2}  \\
- \beta d(x) -\beta \frac{h}{2}
\end{array}\right)
+
{\cal K}*
\left(\begin{array}{c}
\llg{Y^{(\frac{1}{2})}(x)} \\
\vdots \\
\llg{Y^{(S_1-\frac{1}{2})}(x)} \\
\llg{B(x)} \\
\llg{\bar{B}(x)}
\end{array}\right),
\en
where the $(2 S_1+1)\times (2 S_1+1) $ kernel matrix is given by
(\ref{Kernel-x}).

Finally, the largest eigenvalue $\Lambda^{(S_1)}(0)$ of the quantum
transfer matrix $T^{(S_1)}(0)$ (\ref{QTMalterG}) can be written in terms
of the auxiliary functions in analogy to the previous case. We just have to
set $S_2=S_1$ in all expressions (\ref{eingnorm}-\ref{energyS2S1}) and
obtain,
\eq
\llg{\underline{\Lambda}^{(S_1)}(0)}=\beta \left[\psi\left(\frac{2
    S_1+1}{2}\right)-\psi\left(\frac{1}{2}\right)\right]  + \left(K*\llg{B\bar{B}}\right)(0).
\label{einS1S1}
\en
Its contribution to the free energy is given by
\eq
f^{(S_1,S_1)}=\frac{1}{2}\left[\sum_{l=1}^{2 S_1}\frac{1}{l}-\psi\left(\frac{2
    S_1+1}{2}\right)+\psi\left(\frac{1}{2}\right)\right]
-\frac{1}{2\beta}\left(K*\llg{B\bar{B}}\right)(0).
\en

\subsection{$S_1 > S_2$}

For this case, the auxiliary functions as well as the set of non-linear
integral equations are exactly the same as in the previous case $S_2=S_1$. The
only difference consists in the way how the largest eigenvalue
$\Lambda^{(S_2)}(0)$ is expressed in terms of the auxiliary functions.

According to the definition of the $Y$-function, we have an equation similar
to (\ref{auxY}) which relates $\Lambda^{(S_2)}(x)$ and
$Y^{(S_2)}(x)$. This relation can be written explicitly as
\eq
\Lambda^{(S_2)}(x+\frac{\im}{2})\Lambda^{(S_2)}(x-\frac{\im}{2})=
f_{S_2}(x)Y^{(S_2)}(x).
\label{auxYS1S2}
\en

Applying (\ref{FT}) to (\ref{auxYS1S2},\ref{eingnorm}), we obtain
\begin{align}
&\hat{\underline{\Lambda}}^{(S_2)}(k)=\frac{\im k\beta
}{2\cosh{\left[k/2\right]}} \hat{\gamma}(k) + \hat{K}(k)
\hat{Y}^{(S_2)}(k), \\
&\hat{\gamma}(k)=\sum_{\stackrel{l=1}{l>(S_1-S_2)+a}}^{2 S_1}
e^{-|k|(S_2-S_1+l-\frac{1}{2})} - \sum_{\stackrel{l=1}{l<(S_1-S_2)-a}}^{2 S_1}
e^{-|k|(S_1-S_2-l+\frac{1}{2})}- e^{-|k|(\frac{1}{2}+a)},
\end{align}
where $a=0$ when $S_1-S_2$ is an integer number and $a=1/2$ when $S_1-S_2$ is
a half-integer number. Here, we recall that the possibility $l=S_1-S_2$ was
already excluded in the definition of the $\cal L$-operator (\ref{Loperator}).

After performing the inverse Fourier transform and integration over $x$, we
obtain
\eq
\llg{\underline{\Lambda}^{(S_2)}(x)}=\beta \epsilon^{(S_1,S_2)}(x) +
\left(K*\llg{Y^{(S_2)}}\right)(x),
\label{S1S2GS}
\en
with $\epsilon^{(S_1,S_2)}(x)=\int_{-\infty}^{\infty} \frac{\hat{\gamma}(k)
  e^{\im k x} dk}{2\cosh{\left[k/2\right]}}$. At the particular point $x=0$,
$\epsilon^{(S_1,S_2)}(x)$ is given by
\eq
\epsilon^{(S_1,S_2)}(0)=\psi\left(\frac{S_1+S_2+1}{2}\right)-\psi\left(\frac{S_1-S_2+1}{2}\right).
\label{S1S2GS}
\en

Lastly, the contribution to the free energy is written in terms of the
auxiliary function
\bear
f^{(S_2,S_1)}&=&\frac{1}{2}\left[{\sum_{l=1}^{2 S_1}}^{*}\frac{1}{S_2-S_1+l}
  -\psi\left(\frac{S_1+S_2+1}{2}\right)+\psi\left(\frac{S_1-S_2+1}{2}\right)\right]\nonumber\\
&&-\frac{1}{2\beta}\left(K*\llg{Y^{(S_2)}}\right)(0).
\label{energyS1S2}
\ear

It is interesting to compare $\epsilon^{(S_1,S_2)}(x)$ (\ref{S1S2GS}) with the
previous cases (\ref{S2S1GS},\ref{einS1S1}). These expressions can be
naturally written in a unified form as follows
\eq
\varepsilon^{(S_1,S_2)}=\epsilon^{(S_1,S_2)}(0)=\epsilon^{(S_2,S_1)}(0)
=\psi\left(\frac{S_1+S_2+1}{2}\right)-\psi\left(\frac{|S_1-S_2|+1}{2}\right).
\en
According to (\ref{freeEgeral}), the free energy of alternating spin chains
is described by the sum of $\llg{\Lambda^{(S_2)}(0)}$ and
$\llg{\Lambda^{(S_1)}(0)}$. As a result of that, the sum of
$\varepsilon^{(S_1,S_2)}$ and $\varepsilon^{(S_1,S_1)}$ is the ground state
energy of the quantum Hamiltonian ${\cal H}^{(S_1,S_2)}$,
\eq
\epsilon_0=\psi\left(\frac{S_1+S_2+1}{2}\right)-\psi\left(\frac{|S_1-S_2|+1}{2}\right)+
\psi\left(\frac{2S_1+1}{2}\right)-\psi\left(\frac{1}{2}\right),
\en
which is in agreement with the results based on the $2 S$-string hypothesis
for the cases $S_1=1/2,S_2=S$ \cite{ALADIM} and $S_2=S_1=S$ \cite{BABUJIAN}.

The total free energy is the sum of two pieces
$f=f^{(S_2,S_1)}+f^{(S_1,S_1)}$. As we have seen, the term $f^{(S_2,S_1)}$ at
finite temperature can be written as
\eq
f^{(S_2,S_1)}=f_0^{(S_2,S_1)} -\frac{1}{2\beta}
\begin{cases}
\left(K*\llg{B\bar{B}}\right)(0),  & \text{if } S_1 < S_2 \\
\left(K*\llg{B\bar{B}}\right)(0),  &\text{if } S_1 = S_2 \\
\left(K*\llg{Y^{(S_2,S_1)}}\right)(0), & \text{if } S_1 > S_2,
\end{cases}
\en
where $f_0^{(S_2,S_1)}=\frac{1}{2}\left[{\sum^{*}}_{l=1}^{2
    S_1}\frac{1}{S_2-S_1+l} -\varepsilon^{(S_2,S_1)} \right]$. Here we have to
remind that all auxiliary functions, including $B(x)$ and $\bar{B}(x)$,
are different for different cases $S_1< S_2$ and $S_1 \geq S_2$.

We like to mention that results of an analysis similar to that above were
published in \cite{BORTZ} for the study of single Kondo impurities. In the
present study of bulk properties of lattice models, the integral equations
share some algebraic structures with those in \cite{BORTZ}, but have rather
different analytic properties with respect to the driving terms.

\section{Numerical results}
\label{NUMERICAL}
In this section, we present the numerical results obtained for the specific
heat and magnetic susceptibility for the cases $S_1 < S_2, S_1 = S_2$ and $S_1
> S_2$.

\begin{figure}[h]
\begin{center}
\includegraphics[height=10cm,width=10cm]{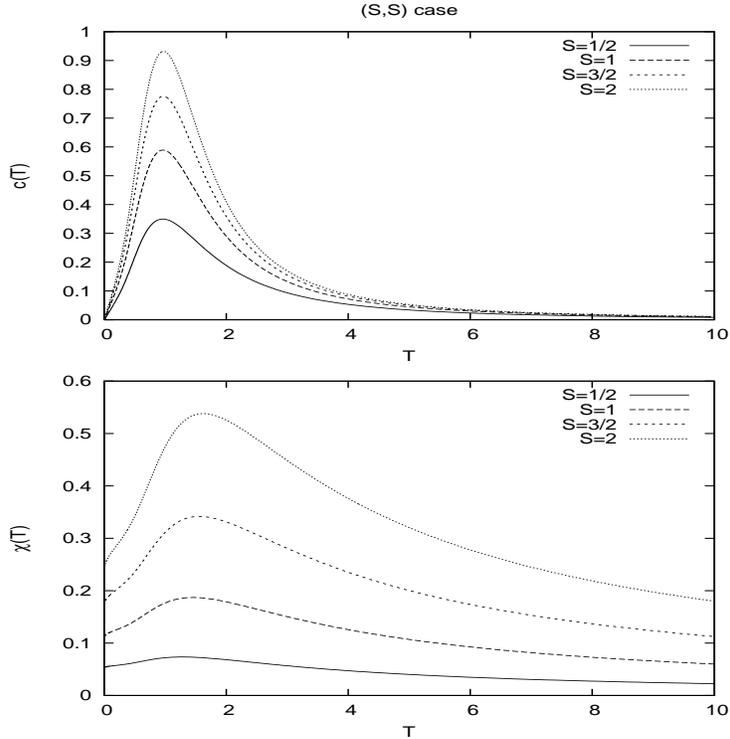}
\caption{Specific heat $c(T)$ and $\chi(T)$ magnetic susceptibility versus
  temperature $T$ for $S=1/2,1,3/2,2$.}
\label{pic1}
\end{center}
\end{figure}

We have solved numerically the non-linear integral equations by iteration. The
convolutions have been calculated in Fourier space using the Fast Fourier
Transform algorithm (FFT). Eventually, we have obtained the free energy as a
function of temperature and magnetic field.

Instead of performing numerical differentiations to obtain the derivatives of
the free energy with respect to temperature and magnetic field, we have
used associated integral equations for the derivatives of the auxiliary
functions. These integral equations arise from the differentiation of the set
of non-linear equations, e.g. with respect to the temperature $T$.

\begin{figure}[h]
\begin{center}
\includegraphics[height=10cm,width=10cm]{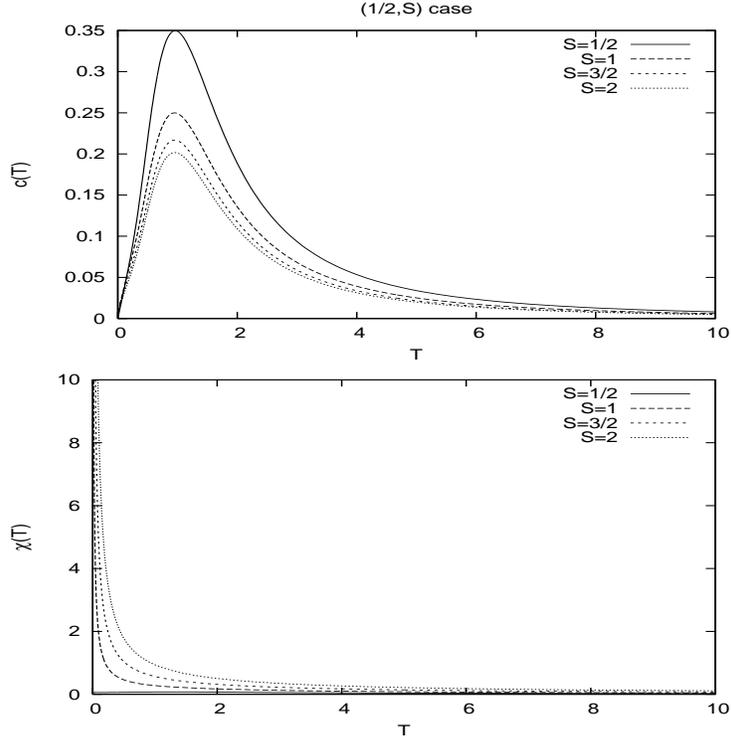}
\caption{Specific heat $c(T)$ and $\chi(T)$ magnetic susceptibility versus temperature $T$ for $S_1=1/2$ and  $S_2=S=1/2,1,3/2,2$.}
\label{pic2}
\end{center}
\end{figure}

Lastly, we have used the relation among the derivatives of the auxiliary
functions reading
\bear
\frac{\partial}{\partial
  T}\llg{B(x)}&=&\frac{b(x)}{1+b(x)}\frac{\partial}{\partial T}\llg{b(x)},\\
\frac{\partial^2}{\partial T^2}\llg{B(x)}&=&\frac{b(x)}{1+b(x)}
\left[\frac{1}{1+b(x)}\left(\frac{\partial}{\partial T}\llg{b(x)}\right)^2+  
\frac{\partial^2}{\partial T^2}\llg{b(x)}\right].
\ear
This way, we obtained for each increment in the order of differentiation a new
set of linear integral equations, where the lower order derivatives appear just
as coefficients.

\begin{figure}[h]
\begin{center}
\includegraphics[height=10cm,width=10cm]{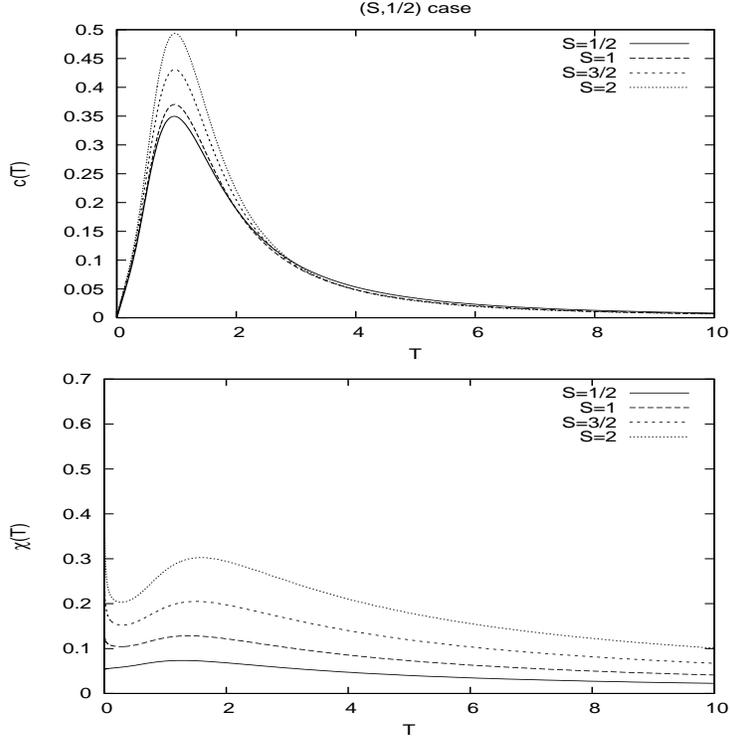}
\caption{Specific heat $c(T)$ and $\chi(T)$ magnetic susceptibility versus
  temperature $T$ for $S_1=S=1/2,1,3/2,2$ and $S_2=1/2$ .}
\label{pic3}
\end{center}
\end{figure}

In Figures \ref{pic1}-\ref{pic3}, we show the specific heat and the magnetic
susceptibility as functions of temperature for the particular cases
$S_1=S_2=S$, $S_1=1/2, S_2=S$ and  $S_1=S, S_2=1/2$ for $S=1/2,1,3/2,2$,
respectively.

The system shows antiferromagnetic behaviour for the first case $S_1=S_2$.  At
low temperature $c(T)$ presents a linear temperature dependence and $\chi(T)$
approaches a finite value. For the case $S_2 > S_1$, we have finite
magnetization $M_f=\frac{S_2-S_1}{2}$ at zero temperature and vanishing
magnetic field $(T=0,h=0^{+})$ in agreement with \cite{DOERFEL}. In the other
limit $(T=0^{+},h=0)$, we have zero magnetization. This is compatible with the
fact that at low temperature and zero magnetic field $\chi(T)$ shows divergent
behaviour. For finite (even small) magnetic field the system becomes polarized
presenting finite magnetization associated with a drop of $\chi(T)$.  In the
last case, $S_1>S_2$, the system behaves as an antiferromagnet. It has zero
magnetization in both limits $(T=0,h=0^{+})$ and $(T=0^{+},h=0)$ in accordance
with \cite{DEVEGA2}.

For the cases $S_2 > S_1$ and $S_2 < S_1$, the models present residual
entropy. The specific values for this quantity can be extracted from low
temperature asymptotic solutions of the non-linear integral equations. The
results are given by $S_{res}=\frac{1}{2}\llg{\left[2(S_2-S_1)+1\right]}$ and
$S_{res}=\frac{1}{2}\llg{\left[\frac{\sin{\frac{\pi
          (2S_2+1)}{2S_1+2}}}{\sin{\frac{\pi }{2S_1+2}}}\right]}$ for
$S_2>S_1$ and $S_2<S_1$ respectively. The latter case was considered in
\cite{DEVEGA2} for $(S_1=1,S_2=1/2)$ using the TBA approach. There, however,
the exact value of the residual entropy was left open due to limitations of
their method.

\section{Thermal current}
\label{thermal}

In this section, we are interested in the thermal Drude weight $D_{th}(T)$ at
finite temperature. We restrict ourselves to the case $S_1=S_2$, where the
thermal current is related to the second conserved charge
(\ref{conserved}) of the transfer matrix.

Specifically, we consider the local conservation of energy in terms of a
continuity equation. This relates the time derivative of the local Hamiltonian
$H_{ii+1}$ to the divergence of the thermal current $j^{E}$, $\dot{H}=-\nabla
j^{E}$. Here, the local term $H_{ii+1}$ stands for
\eq
H_{ii+1}=P_{i,i+1}\frac{\partial}{\partial \lambda}{\cal L}_{i,i+1}^{(S_1,S_1)}(\lambda)\Big|_{\lambda=0}, ~~ {\cal
  H}=\sum_{i=1}^{L}H_{ii+1}.
\en
As the time derivative leads to the commutator with the Hamiltonian, we
obtain
\eq
\dot{H}_{i,i+1}=\im \left[{\cal H}, H_{i,i+1}(t) \right]=-\im \left(
  j_{i+1}^E(t) - j_i^E(t) \right),
\en
where the local energy current $j_i^E$ is given by
\eq
j_i^E=\im \left[ H_{i-1i}, H_{ii+1} \right],
\en
and the total thermal current is ${\cal J}_E=\sum_{i=1}^L j_i^{E}$.

On the other hand, just by comparing the expression for ${\cal J}_E$ and the
second logarithmic derivative of the transfer matrix ${\cal J}^{(2)}$, we
obtain
\eq
{\cal J}_E={\cal J}^{(2)} + \im  \frac{L}{2}\frac{\partial^2}{\partial \lambda^2} {\zeta}_{S_1,S_1}(\lambda)\Big|_{\lambda=0}.
\en

The transport coefficients are determined from the Kubo formula \cite{KUBO} in
terms of the expectation value of the thermal current ${\cal J}_E$, such that
\cite{ZOTOS,SAKAI}
\eq
D_{th}(T)=\beta^2 \left\langle {\cal J}_{E}^2 \right\rangle.
\en

In order to calculate the expectation value $\left\langle {\cal J}_{E}^2
\right\rangle$, we introduce a new partition function $\bar{Z}$ as,
\eq
\bar{Z}=\tr{\left[\exp{\left(-\beta {\cal H}-\lambda_n {\cal
          J}^{(n)}\right)}\right]}.
\en
In this way, we obtain the expectation values of ${\cal J}^{(2)}$ through the
logarithmic derivative of $\bar{Z}$,
\eq
\left(\frac{\partial}{\partial \lambda_2}\right)^2
\llg{\bar{Z}}\Big|_{\lambda_2=0}=\left\langle {\cal J}_{E}^2 \right\rangle,
\en
where we used the fact that the expectation value of the thermal current in
thermodynamical equilibrium is zero $\left\langle {\cal J}_{E}
\right\rangle=0$.

To compute the partition function $\bar{Z}$, we consider the procedure
developed in \cite{SAKAI}. We rewrite the partition function $\bar{Z}$ in
terms of the row-to-row transfer matrix such that
\bear
\bar{Z}&=&\lim_{N\rightarrow \infty}\tr{\left[\exp{\left( T(u_1)\dots T(u_N)
        T(0)^{-N} \right)}\right]}, \nonumber\\
&=&\tr{\left[\exp{\left( \lim_{N\rightarrow \infty} \sum_{l=1}^{N}\{
        \llg{T(u_l)}- \llg{T(0)} \} \right)}\right]}.
\ear
The numbers $u_1,\dots,u_N$ are chosen in such a way that the following
relation is satisfied, 
\eq
\lim_{N\rightarrow \infty}\sum_{l=1}^{N}\{ \llg{T(u_l)}- \llg{T(0)} \}=-\beta
\frac{\partial}{\partial x}\llg{T(x)}\Big|_{x=0}+\lambda_n
\im^{n-1}\frac{\partial^n}{\partial x^n}\llg{T(x)}\Big|_{x=0}.
\en

We can proceed analogously to section \ref{QTM} and introduce a quantum
transfer matrix associated to the partition function $\bar{Z}$. Instead of the
staggered vertex model with alternation in vertical direction between
$T(-\tau)$ and $\overline{T}(-\tau)$, we have now $N$ different terms of the
form $T(0)^{-1}T(u_l)$ for $l=1,\dots,N$. As $T(0)^{-1}=\overline{T}(0)/{\cal N}$, we can
write $T(0)^{-1}=[(2 S_1)!]^{-2 L} T(-\rho)$. So, we have the alternation of
$T(-\rho)$ and $T(u_l)$ which is a special case of the previous sections. 

Therefore, we can proceed along the same lines as before which is equivalent
to substitute $\phi_{+}(x) \rightarrow \prod_{l=1}^{N} \phi_{l}(x)$ and
$\phi_{-}(x) \rightarrow \prod_{l=1}^{N} \phi_{0}(x)$ where $\phi_{l}(x) =
x-\im u_l$ and $\phi_{0}(x) = x$.

In this way, the partition function can be written in the thermodynamical
limit in terms of the largest eigenvalue,
\eq
\lim_{L\rightarrow \infty}\frac{1}{L} \llg{\bar{Z}}=\llg{\Lambda(0)},
\en
which is written as
\eq
\llg{\Lambda}(0)=(-\beta + \lambda_n \frac{\partial^{n-1}}{\partial x^{n-1}} )
{\cal E}(x)\Big|_{x=0} + \left(K*\llg{B\bar{B}}\right)(0),
\en
where ${\cal E}(x)=\epsilon^{(S_1,S_1)}(x)$.

The auxiliary functions $B$ and $\bar{B}$ satisfy the following set of
non-linear integral equations
\eq
\left(
\begin{array}{c}
\llg{y^{(\frac{1}{2})}(x)} \\
\vdots \\
\llg{y^{(S_1-\frac{1}{2})}(x)} \\
\llg{b(x)} \\
\llg{\bar{b}(x)}
\end{array}\right)=
\left(\begin{array}{c}
0 \\
\vdots \\
0 \\
(-\beta+ \lambda_n \frac{\partial^{n-1}}{\partial x^{n-1}} ) d(x)  \\
(-\beta+ \lambda_n \frac{\partial^{n-1}}{\partial x^{n-1}} ) d(x)  
\end{array}\right)
+
{\cal K}*
\left(\begin{array}{c}
\llg{Y^{(\frac{1}{2})}(x)} \\
\vdots \\
\llg{Y^{(S_1-\frac{1}{2})}(x)} \\
\llg{B(x)} \\
\llg{\bar{B}(x)}
\end{array}\right).
\en

Therefore, the thermal Drude weight is given by,
\eq
D_{th}(T)=\beta^2 \left\langle {{\cal J}^{(2)}}^2 \right\rangle=\beta^2
\left(\frac{\partial}{\partial
    \lambda_2}\right)^2\llg{\Lambda}(0)\Big|_{\lambda_2=0}.
\en

\begin{figure}[h]
\begin{center}
\includegraphics[height=10cm,width=10cm]{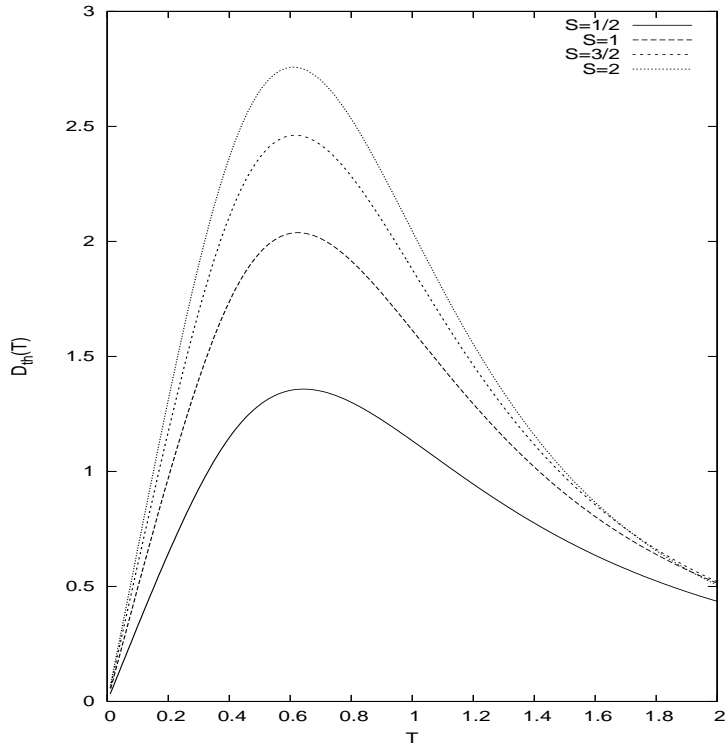}
\caption{Thermal Drude weight $D_{th}(T)$ as function of temperature for
  $S=1/2,1,3/2,2$.}
\label{pic4}
\end{center}
\end{figure}

In Figure \ref{pic4}, we show the thermal Drude weight as function of the
temperature for $S_1=S_2=S$. It exhibits a linear behaviour at low
temperatures and is proportional to the central charge $c=\frac{3 S}{S+1}$. This is in agreement with the
spin-$1/2$ case \cite{SAKAI}.

Before closing this section, we would like to mention that in the general case
$(S_1,S_2)$ the thermal current does not look like a conserved current. In
this case, we cannot provide exact results for the Drude weight. Nevertheless,
we are able to provide an exact description of the second logarithmic
derivative of the transfer matrix. However, the physical interpretation of
this quantity has eluded us so far.

\section{Conclusion}
\label{CONCLUSION}
In this paper we managed to construct the quantum transfer matrix for the case
of non-isomorphic auxiliary and quantum spaces of interacting spins. We
considered explicitly the generic $(S_1,S_2)$ case of alternating spin chains
and obtained a finite set of non-linear integral equations. These equations
were solved numerically for the cases $S_1 < S_2$ and $S_1 \geq S_2$. In this
way, we obtained the specific heat and the magnetic susceptibility as
functions of temperature. For the particular case $S_1=S_2$, we also provided
results for the thermal Drude weight at finite temperature.

The system behaves antiferromagnetically for $S_1 \geq S_2$ and presents finite magnetization in the remaining case $S_1 < S_2$. Interestingly, for all $S_1 \neq S_2$ we have residual entropy at zero temperature which we were able to evaluate exactly. Recently, systems with finite entropy at $T=0$ attracted interest regarding efficient cooling procedures \cite{HONECKER}.

We expect that our results may be interesting for the study of generic
mixed spin chains \cite{MIX}.  Another interesting issue deserving
investigation is the physical interpretation of the second conserved charge
for the generic case $(S_1,S_2)$ and its implications on transport properties.

\section*{Acknowledgments}
The authors thank DFG (Deutsche Forschungsgemeinschaft) for financial support
and G.A.P. Ribeiro thanks J. Damerau for many useful discussions.

\end{document}